# Dimeric Drug Polymeric Micelles with Acid-Active Tumor Targeting and FRET-indicated Drug Release

Xing Guo, [†,‡,⊥] Lin Wang, [†,‡,⊥] Kayla Duval, [†] Jing Fan, Shaobing Zhou, [‡,*] and Zi Chen[†,*]

[†]Thayer School of Engineering, Dartmouth College, Hanover, New Hampshire 03755, USA

[‡]Key Laboratory of Advanced Technologies of Materials, Ministry of Education, School of Materials Science and Engineering, Chengdu 610031, China

**ABSTRACT:** Trans-activating transcriptional activator (TAT), a cell-penetrating peptide, has been extensively used for facilitating cellular uptake and nuclear targeting of drug delivery systems. However, the positively charged TAT peptide usually strongly interacts with serum components and undergoes substantial phagocytosis by the reticuloendothelial system, causing a short blood circulation *in vivo*. In this work, an acid-active tumor targeting nanoplatform DA-TAT-PECL was developed to effectively inhibit the nonspecific interactions of TAT in the bloodstream. 2,3-dimethylmaleic anhydride (DA) was first used to convert the TAT's amines to carboxylic acid, the resulting DA-TAT was further conjugated to poly(ethylene glycol)-poly(ε-caprolactone) (PEG-PCL, PECL) to get DA-TAT-PECL. After self-assembly into polymeric micelles, they were capable of circulating in the physiological condition for a long time and promoting cell penetration upon accumulation at the tumor site and de-shielding the DA group. Moreover, camptothecin (CPT) was used as the anticancer drug and modified into a dimer $(CPT)_2$-ss-Mal, in which two CPT molecules were connected by a reduction-labile maleimide thioether bond. The Förster resonance energy transfer (FRET) signal between CPT and maleimide thioether bond was monitored to visualize the drug release process and effective targeted delivery of antitumor drugs was demonstrated. This pH/reduction dual-responsive micelle system provides a new platform for high fidelity cancer therapy.

## INTRODUCTION

Cancer is the leading cause of death in the world today, with over 1,600,000 new cancer cases diagnosed and over 580,000 cancer deaths in the U.S. in 2015, according to the American Cancer Society.[1] Although traditional chemotherapy remains to be a prior method against cancer, chemotherapeutics usually fail due to the poor water solubility, high toxicity and low bioavailability. To solve these problems, nanotechnology-mediated drug delivery system (NDDS) has emerged. Nanoparticles (NPs), especially those that are made from biodegradable and biocompatible polymers, have been widely utilized as NDDS for cancer therapy.[2-4] Using nanotechnology, it is possible to achieve improved delivery of poorly water-soluble drugs, targeted delivery of drugs in a cell- or tissue-specific manner, co-delivery of two or more drugs for combination therapy, etc.[5] Hence with all these advantages, NPs can be an outstanding candidate for loading chemotherapeutics.

A key goal of NPs is to quickly deliver their payloads into desired cells. Cell-penetrating peptides (CPPs), highly cationic peptides usually rich in arginine and lysine amino acids,[6] have been extensively explored for decorating NPs to promote their cellular uptake. This is ascribed to the high affinity of positively charged CPPs for negatively charged cell membranes. Some sequences, such as the protein transduction domain from the HIV trans-activating transcriptional activator (TAT) peptide (CYGRKKRRQRRR),[7] can be recognized by the nuclear pore complexes (NPCs),[8] and thus actively transport cargos from the cytosol into the cell nucleus. Therefore, TAT peptide is one of the most thoroughly studied and extensively used CPPs for intracellular and nuclear delivery.

However, TAT-modified NPs with a positively charged surface undergo substantial phagocytosis by the reticular endothelin system (RES) and interact strongly with serum components, causing severe aggregation and a short blood circulation half-life.[9] Moreover, TAT peptides are internalized by both normal cells and tumor cells due to their lack of tumor targeting property. In order to effectively use TAT-modified NPs for cancer treatment, a new method that can prevent aggregation and increase both the half-life time in the blood circulation and the tumor targeting capability needs to be developed. Previous studies have proven that the NPs containing amides with $\beta$-carboxylic acid groups are acid labile; as a result, they are negatively charged under a neutral condition while becoming positively charged in an acidic environment, followed by a selective internalization by tumor cells instead of normal cells.[10,11] Therefore, the TAT's positive charge can be shielded by converting their amines to $\beta$-carboxylic amides when triggered by an environmental pH change, making the modified TAT exhibit pH-dependent hydrolysis and achieving a negative-to-positive charge reversal.

Apart from high stability and tumor targeting, an ideal drug delivery vehicle ought to precisely control the release of cargo as well. In order to achieve site-specific drug delivery, strategies have been developed by introducing stimuli responsiveness into NPs, making them respond to modest environmental changes, which in turn leads to alterations in their structures or chemistry subsequently inducing the release of drug.[12-14] Owing to the differential glutathione tripeptide (γ-glutamyl-cysteinyl-glycine, GSH) level between the intracellular compartment (2~10 mM) and the extracellular microenvironment (2~20 μM),[15] the significant redox potential difference can be used as a unique trigger for designing NDDS with programmed release of chemotherapeutics intracellularly.[16,17] Compared to the disulfide bond that has been widely applied for constructing reduction-responsive systems, another bioreducible bond, dithiomaleimides, produced by reaction of 2,3-dibromomaleimide and thiols,[18,19] possess better thermal- and photo-stability than disulfide bonds.[20] Furthermore, the strong green fluorescence of dithiomaleimides can be taken advantage of for tracking the NDDS.[21,22]

Here, we develop an acid-triggered tumor targeting polymeric micelle system loaded with a reduction-sensitive camptothecin (CPT) dimer, which is capable of self-reporting drug release via FRET. Polymeric micelles that possess a unique core/shell architecture and nanoscale size are widely used as drug carriers for loading anticancer drugs. In this nanoplatform, polyethylene (PEG) and poly(ε-caprolactone) (PCL) were used as the hydrophilic segment and hydrophobic segment, respectively. To allow the micelles to be internalized by cancer cells, TAT peptide was conjugated to the PEG segment. Further, the lysine and glutamine residues' amines of TAT were amidized by 2,3-dimethylmaleic anhydride (DA) to prevent nonspecific interactions *in vivo* resulting from the cationic nature of the amines (**Figure 1**).

Next, the anticancer drug, CPT was modified into a dimer $(CPT)_2$-ss-Mal, whereby two CPT molecules were connected by a reduction-labile maleimide thioether bond. The dimer featured an emission peak at 550 nm (dithiomaleimide emission wavelength, green fluorescence) rather than at 438 nm (CPT emission wavelength, blue fluorescence) when excited at 370 nm (CPT excitation wavelength), indicating the existence of the FRET phenomenon between CPT (donor) and maleimide thioether bond (acceptor). However, in the presence of a reducing environment, the maleimide thioether bond can be cleaved, subsequently releasing CPT via cyclization reaction. As a result, the FRET effect was abolished accordingly with the transition of emission peak to 438 nm and the recovery of blue fluorescence when excited at 370 nm. By monitoring the change of the FRET signal on the CPT dimer-loaded micelles in real time, we can visualize and quantify the drug release process (**Figure 1**).

The $(CPT)_2$-ss-Mal-loaded DA-TAT-PECL micelles with a negatively charged surface possessed high stability in blood circulation due to the electrostatic effect. Once they accumulated at the acidic tumor tissues through highly permeable blood vessels via the enhanced permeability and retention (EPR) effect,[23] the amides in the DA-TAT quickly hydrolyzed, fully restoring TAT's functions and facilitating the cellular entry. Afterwards, CPT were released from $(CPT)_2$-ss-Mal triggered by the intracellular GSH with FRET-indicated signal change. This development of combined acid-active targeting with reducible drug dimer and FRET-based self-reporting drug release provides a novel approach for efficient cancer therapy.

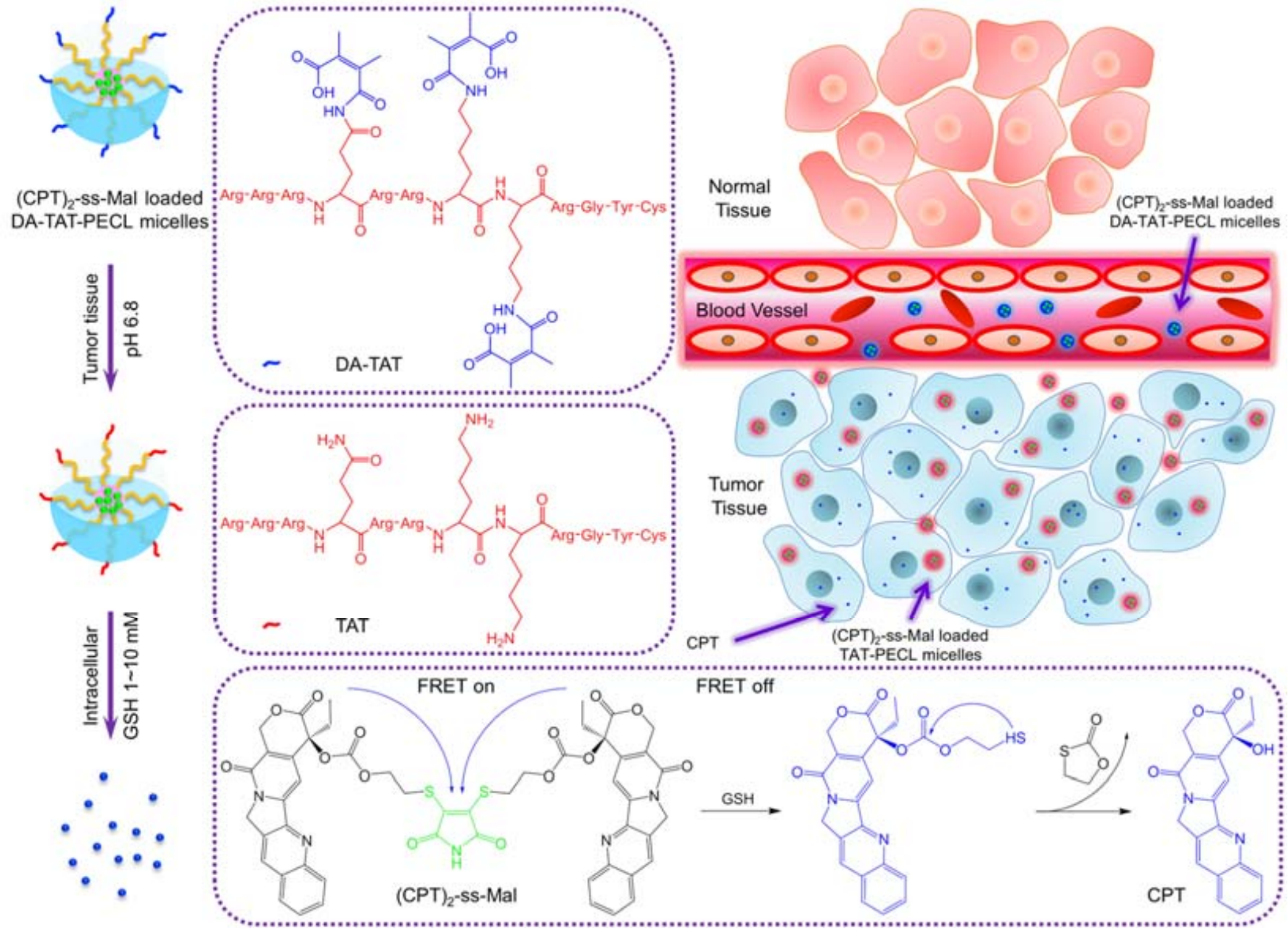

**Figure 1**. Schematic illustration of acid-triggered tumor targeting DA-TAT-PECL nanoplatform loading with reduction-sensitive $(CPT)_2$-ss-Mal. Shielding of the cationic charges of TAT with DA suppresses its nonspecific interactions in blood circulation, while regenerating the original TAT in weakly acidic tumor tissue (pH 6.8) for cellular

internalization or in more acidic endo/lysosomes (pH 4~5) for endo/lysosomal escape. Release of CPT from $(CPT)_2$-ss-Mal is selectively triggered by the high concentration of intracellular GSH (2~10 mM), and the concomitant change of FRET signal can be used to report the drug release process.

## RESULTS AND DISCUSSION

**Synthesis and Characterization of DA-TAT-PECL Polymer and $(CPT)_2$-ss-Mal Drug Dimer.** DA-TAT-PECL was synthesized by amidization of the TAT lysine and glutamine residues' amines using 2,3-dimethylmaleic anhydride (DA), followed by cojugating DA-TAT to Mal-PECL which was synthesized through ring-opening polymerization (ROP) of ε-caprolactone initiated by Mal-PEG-OH in the presence of stannous octoate ($Sn(Oct)_2$) (**Figure 2A**). $(CPT)_2$-ss-Mal was synthesized by activation of CPT with triphosgene and then reaction with 2-hydroxyethyl disulfide to get disulfide linked-$(CPT)_2$-ss, which was further conjugated to 2,3-dibromomaleimide in the presence of tris(2-carboxyethyl)phosphine (TCEP) to yield $(CPT)_2$-ss-Mal (**Figure 2B**).

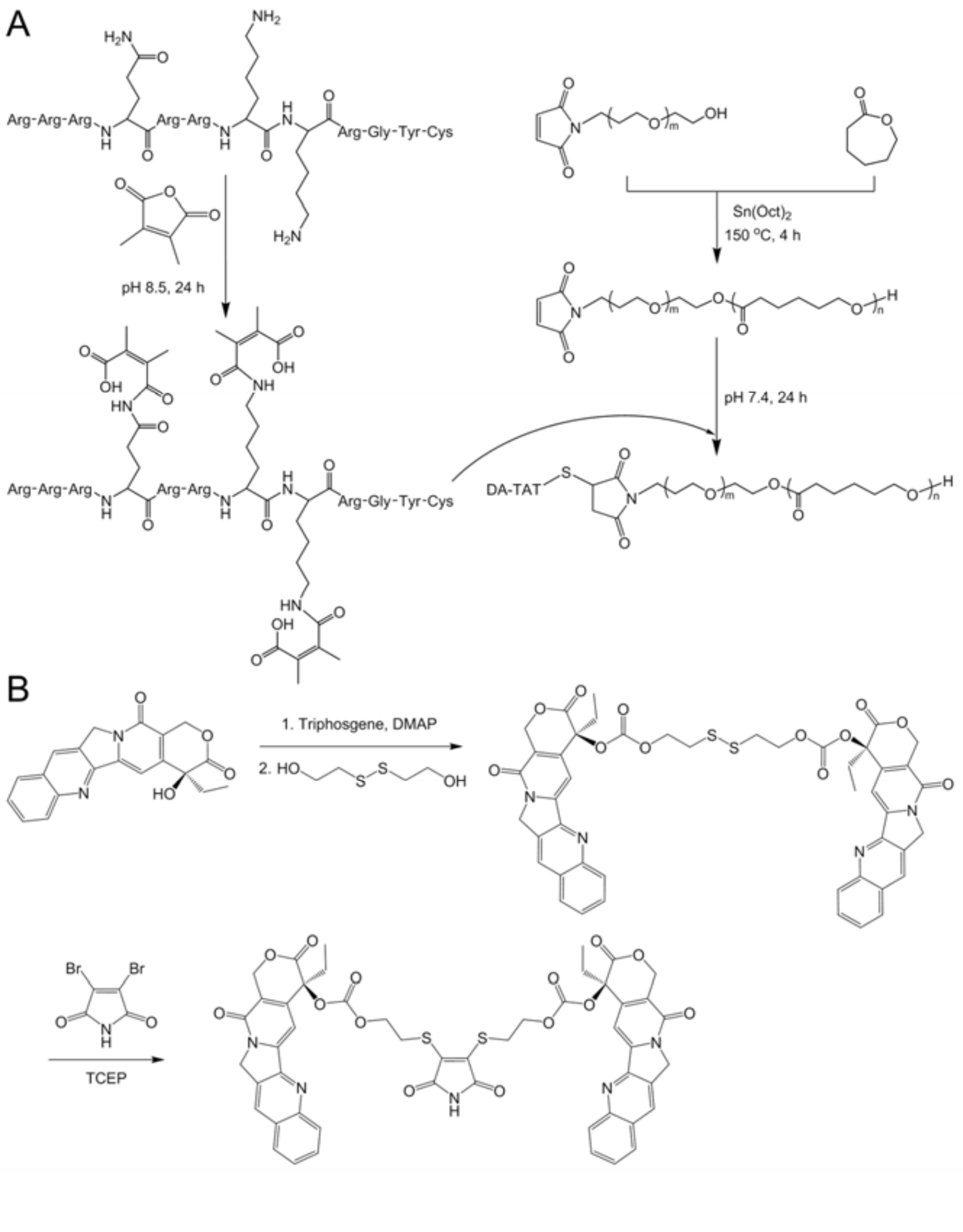

**Figure 2.** Synthesis of (A) DA-TAT-PECL polymer and (B) $(CPT)_2$-ss-Mal drug dimer.

According to the ratio of integral areas of different peaks in $^1$H NMR, the average degree of polymerization of PCL segment can be calculated as 59 (**Figure 3A**). Correspondingly, the molecular weight of Mal-PECL is 11700. As for the TAT-PECL, the characteristic resonance of maleimide at 6.68 ppm disappeared, indicating the successful conjugation of TAT to Mal-PECL (**Figure 3B**).[24] Besides, the appearance of the peak at 1.84 ppm in $^1$H NMR of DA-TAT-PECL indicated the complete amidation of TAT amines by DA (**Figure 3C**).[10] The signals of hydrogen protons adjacent to difsulfide bond in $(CPT)_2$-ss appeared at 2.83 ppm and 4.02 ppm, while transferred to 3.08 ppm and 4.41 ppm after reaction with 2,3-dibromomaleimide (**Figure 3D** and **3E**). This change in chemical shift is caused by the electron-withdrawing inductive effect of maleimide.[25]

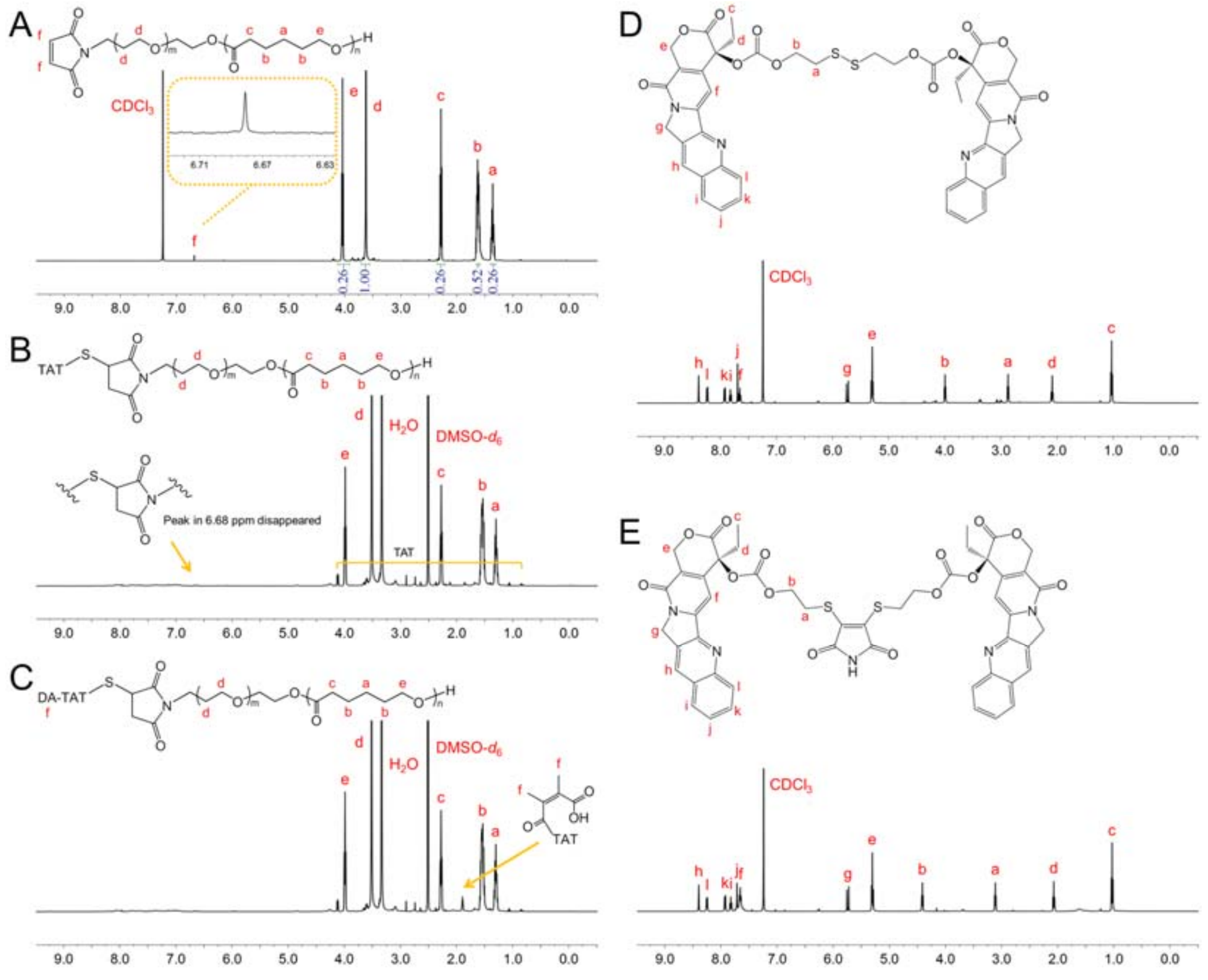

**Figure 3.** $^1$H NMR spectra of (A) Mal-PECL, (B) TAT-PECL, (C) DA-TAT-PECL, (D) $(CPT)_2$-ss and (E) $(CPT)_2$-ss-Mal in $CDCl_3$ or in DMSO-$d_6$.

**Characterization of Micelles and pH/Reduction Dual-Responsiveness.** The average size of DA-TAT-PECL micelles was 43.9 nm with low polydispersity index (PDI) of 0.135, indicating a narrow distribution of diameters (**Figure 4A**). The TEM image showed that the DA-TAT-PECL micelles exhibited a spherical shape and a nearly uniform distribution. The diameter of TEM image was consistent with that detected by DLS (**Figure 4A**).

Shielding of the cationic charges of TAT with DA was supposed to inhibit the nonspecific interaction in blood circulation, as the negatively charged components in the surface of blood vessels and the surface of cells would repel negatively charged nanoparticles.[26] To validate this, the average size of TAT-PECL and DA-TAT-PECL micelles in the presence of 0.5 mg/mL bovine serum albumin (BSA) was measured by DLS at different time points. There's no significant size change in cationic charge-shielded DA-TAT-PECL micelles after 48 hours (h)' BSA incubation (**Figure 4B**). In contrast, the cationic charge-unshielded TAT-PECL micelles increased their size rapidly in a short time. The result indicated that the DA-decorated micelles could effectively resist protein adsorption as expected.

Amides with $\beta$-carboxylic acid groups are acid labile,[27] in which DA-TAT-PECL micelles ought to be negatively charged in neutral condition while reversing to be positively charged triggered by acid due to the recovery of amines upon amides hydrolysis.[11] As shown in **Figure 4C**, DA-TAT-PECL micelles were negatively charged at the beginning regardless of the environmental pH value. However, the ζ-potential of these micelles quickly increased and became positive within 5 minutes at pH 6.8, while it remained negative within 120 minutes at pH 7.4. Considering the negative charge of cell membranes, the charge-reversal behavior of DA-TAT-PECL micelles at tumor extracellular pH would enhance their internalization by tumor cells.[28] Besides, the exposure of TAT peptide with the cleavage of amides in DA-TAT-PECL micelles would further facilitate the cellular uptake, followed by endo/lysosomal escape in more acidic endo/lysosomal compartment.[29] In contrast to DA-TAT-PECL micelles, TAT-PECL micelles always stayed positively charged with ζ-potentials around 10 mV at any environmental pH value (**Figure 4D**).

To confirm the reduction-responsive FRET inactivation, the fluorescence properties of $(CPT)_2$-ss-Mal in the presence of GSH were evaluted by fluorescence spectrometer and digital photos. When excited at 370 nm (CPT excitation wavelength), the fluorescence spectrum of the solution showed an emission peak at 550 nm (dithiomaleimide emission wavelength) instead of 438 nm (CPT emission wavelength), indicating the existence of the FRET phenomenon between CPT and maleimide thioether bond (**Figure 4E**). However, a decrease in dithiomaleimide fluorescence and an increase in CPT fluorescence were observed with GSH incubation, while the peak at 550 nm even disappeared 60 minutes later, indicating a complete cleavage of maleimide thioether bond. This was also reflected in a significant color change of the solution from green to blue under UV irradiation (365 nm) (**Figure 4E**, inset). According to these results, it can be inferred that $(CPT)_2$-ss-Mal rapidly released CPT in a reducing environment accompanied by change in the FRET signal.

Owing to the π-π interaction of the planar pentacyclic aromatic ring structure, CPT molecules self-aggregated to micrometer-size large particles when encapsulated into micelles, leading to a very low drug loading (usually less than 5%).[30] By covalently linking two CPT molecules into a dimer, formation of large aggregates was prevented due to the inhibition of long-range-ordered drug molecule packing. Consequently, high drug-loading content (LC) and encapsulation efficiency (EE) would be expected by loading the CPT dimers into micelles. To confirm this, the size and drug content of $(CPT)_2$-ss-Mal-loaded DA-TAT-PECL micelles were measured by DLS and UV-vis, respectively. The results demonstrated that these particles were a little bigger than the blank micelles with a diameter of 51.6 nm. Besides, the LC and EE were as high as 42% and 84%, respectively. Next, we detected the *in vitro* CPT release from the $(CPT)_2$-ss-Mal-loaded DA-TAT-PECL micelles in the absence of GSH and in the presence of 10 μM or 10 mM GSH. Negligible CPT release was observed over 48 h without GSH treatment or treated with 10 μM GSH (**Figure 4F**). In comparison, a rapid release of CPT was immediately turned on in the presecne of 10 mM GSH, in which over 90% of CPT was released from the micelles within the experimental time. Therefore, the ideal drug release behavior of this system was confirmed. The drug delivery system is shown to be stable in the physiological condition (~10 μM GSH), while resopnding to intracellular reducing environment (~10 mM GSH) and subsequently releasing the cargo.

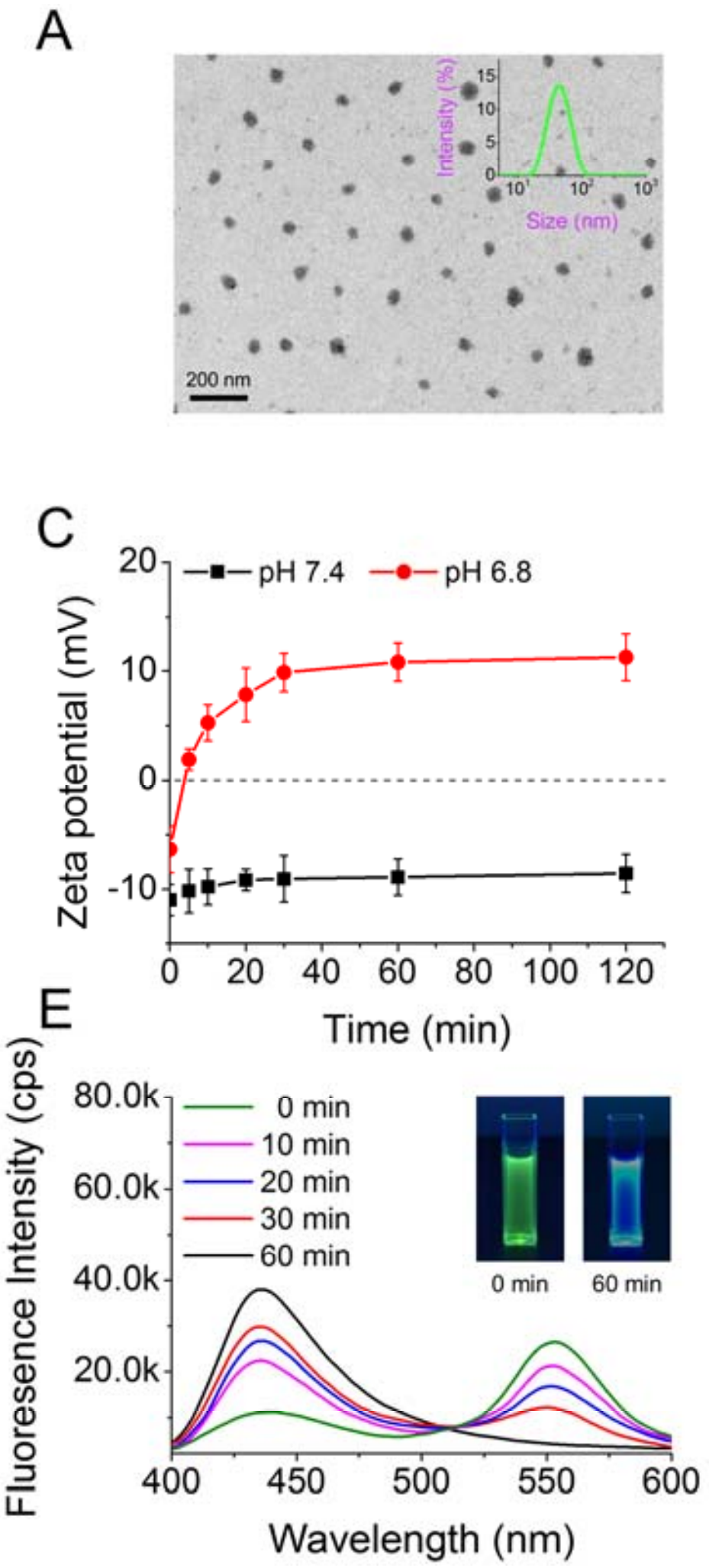

**Figure 4.** (A) Size distribution detected by DLS and TEM image of DA-TAT-PECL micelles. (B) Size change in TAT-PECL and DA-TAT-PECL micelles after incubation with 0.5 mg/mL bovine serum albumin (BSA). Zeta-potential change in DA-TAT-PECL (C) and TAT-PECL (D) micelles at pH 7.4 and pH 6.8. (E) Fluorescence spectra of $(CPT)_2$-ss-Mal in the presence of 10 mM GSH at different time points. Excitation wavelength was set at 370 nm. Inset: pictures of $(CPT)_2$-ss-Mal in the presence of 10 mM GSH under UV irradiation (365 nm) for 0 minutes (left) and 60 minutes (right). (F) Percent CPT released from the $(CPT)_2$-ss-Mal-loaded DA-TAT-PECL micelles in the absence or presence of 10 μM and 10 Mm GSH.

**Cell Cytocompatibility.** To study the cytocompatibility of these micelles as potential drug delivery systems, we performed alamarBlue assay and live/dead staining. MDA-MB-231 cell viability was over 90% with incubation of different concentrations of PECL, TAT-PECL and DA-TAT-PECL micelles for 48 h (**Figure 5A**). The fluorescence images showed that there were no dead cells unless treated with a high concentration of micelles at 500 μg/mL (**Figure 5B**). Taken together, we believe these vehicles possess excellent cytocompatibility and potential application as drug carreris.

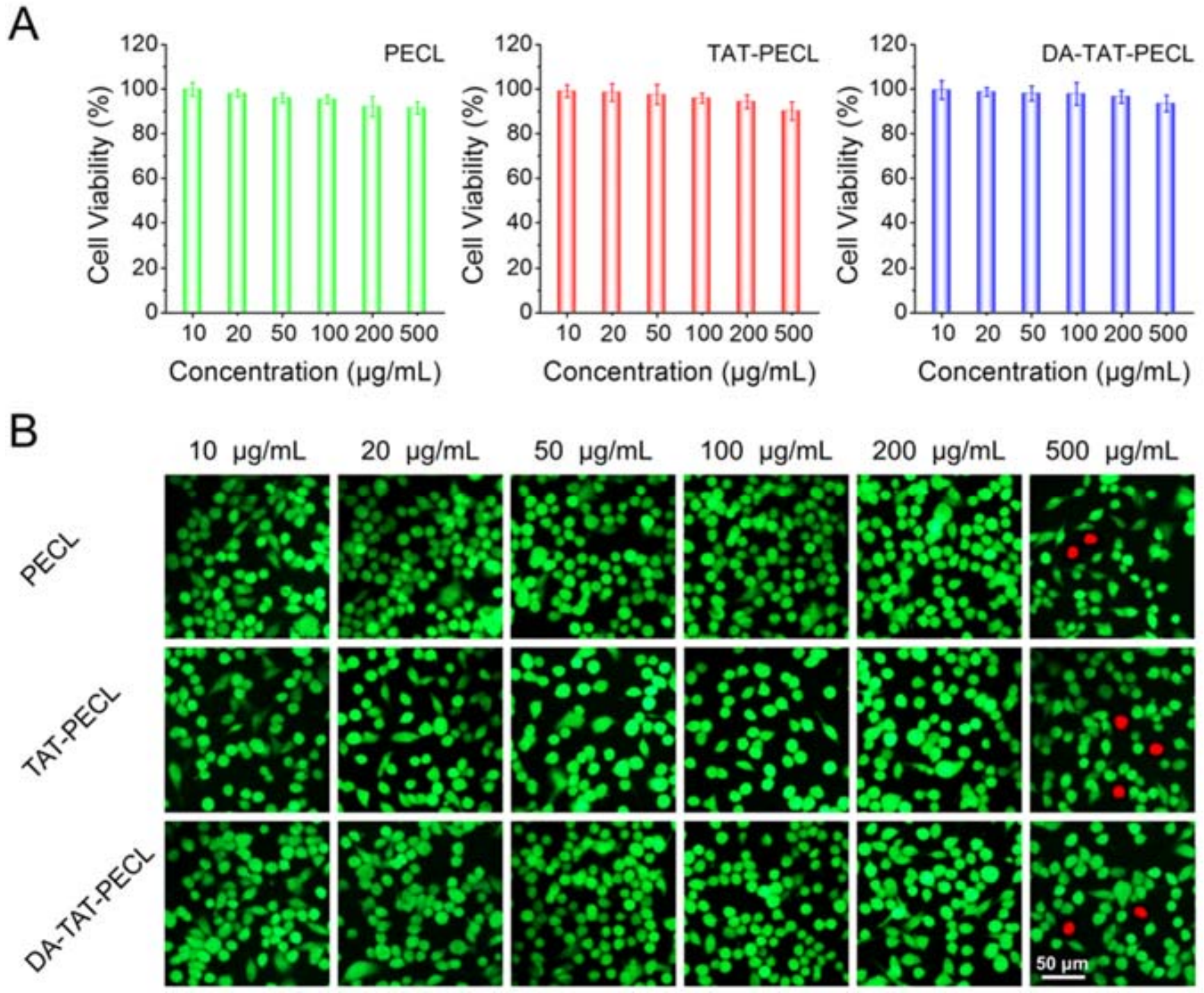

**Figure 5.** (A) Viability of MDA-MB-231 cells by alamarBlue assay after treating with different concentrations of PECL, TAT-PECL and DA-TAT-PECL micelles for 48 h (n=5). (B) Fluorescence images showing the viability of MDA-MB-231 cells by calcein AM and propidium iodide double staining after treating with different concentrations of PECL, TAT-PECL and DA-TAT-PECL for 48 h. The live cells were stained green while the dead cells were stained red.

**Intracellular FRET Inactivation.** The reducing intracellular compartment is expected to trigger the cleavage of maleimide thioether bond in $(CPT)_2$-ss-Mal once it is internalized by cells, followed by the release of CPT. Interestingly, this prcoess can be self-reporting via FRET signal. To demonstrate this, we treated MDA-MB-231 cells with 1 mM $(CPT)_2$-ss-Mal and observed the real-time fluorescence images using 365 nm as excitation wavelength. As shown in **Figure 6A**, the cells overall showed green fluorescence within 30 minutes, indicating a cellular uptake process without CPT leakage. After 45 minutes of incubation, the blue fluorescence appeared in the cells and increased in intensity as time passed by. Finally, all the cells showed blue fluorescence with the disappearance of green fluorescence at 90 minutes. The change in FRET signal indicated the *in situ* release of CPT from $(CPT)_2$-ss-Mal between 45 minutes to 90 minutes. It is also worth mentioning that the intracellular GSH-induced FRET inactivation process took as long as that measured by the fluorescence spectrometer (**Figure 4E**). On the basis of these images, we were able to confirm that the drug release occurred in response to the intracellular GSH.

**pH-Dependent Cellular Uptake.** Compared to the negatively charged NPs, vehicles with a positively charged surface tend to possess a higher rate of cellular uptake.[31] The charge reversal of DA-TAT-PECL micelles in weak acidic condition was confirmed (**Figure 4C**), consequently, they ought to be more efficiently internalized by cells once reversing to positive charge and exposure of TAT peptide. To observe the pH-dependent cellular uptake, MDA-MB-231 cells were treated with $(CPT)_2$-ss-Mal-loaded PECL, TAT-PECL and DA-TAT-PECL micelles at pH 7.4 and pH 6.8 for 2 h, and imaged by the fluorescence microscope. As shown in **Figure 6B**, weak green fluorescence was observed in DA-TAT-PECL group at pH 7.4, indicating a slow cellular uptake of these micelles. However, they were remarkably internalized at pH 6.8 and completely released CPT within 2 h, in which blue fluorescence as strong as the TAT-PECL group was displayed. The cellular uptake of the neutral PECL micelles without TAT decoration was irrelevant to environmental pH, where similar fluorescence intensity was observed in the cells at pH 7.4 and 6.8. The flow cytometry analysis further proved that the endocytosis of DA-TAT-PECL micelles by MDA-MB-231 cells was pH dependent. After 2 h of incubation, the CPT fluorescence in the DA-TAT-PECL group was significantly enhanced at pH 6.8 compared to that at pH 7.4 (**Figure 6C**). There was no obvious change of the CPT fluorescence in cells treated with either PECL micelles or TAT-PECL micelles at pH 7.4 and pH 6.8.

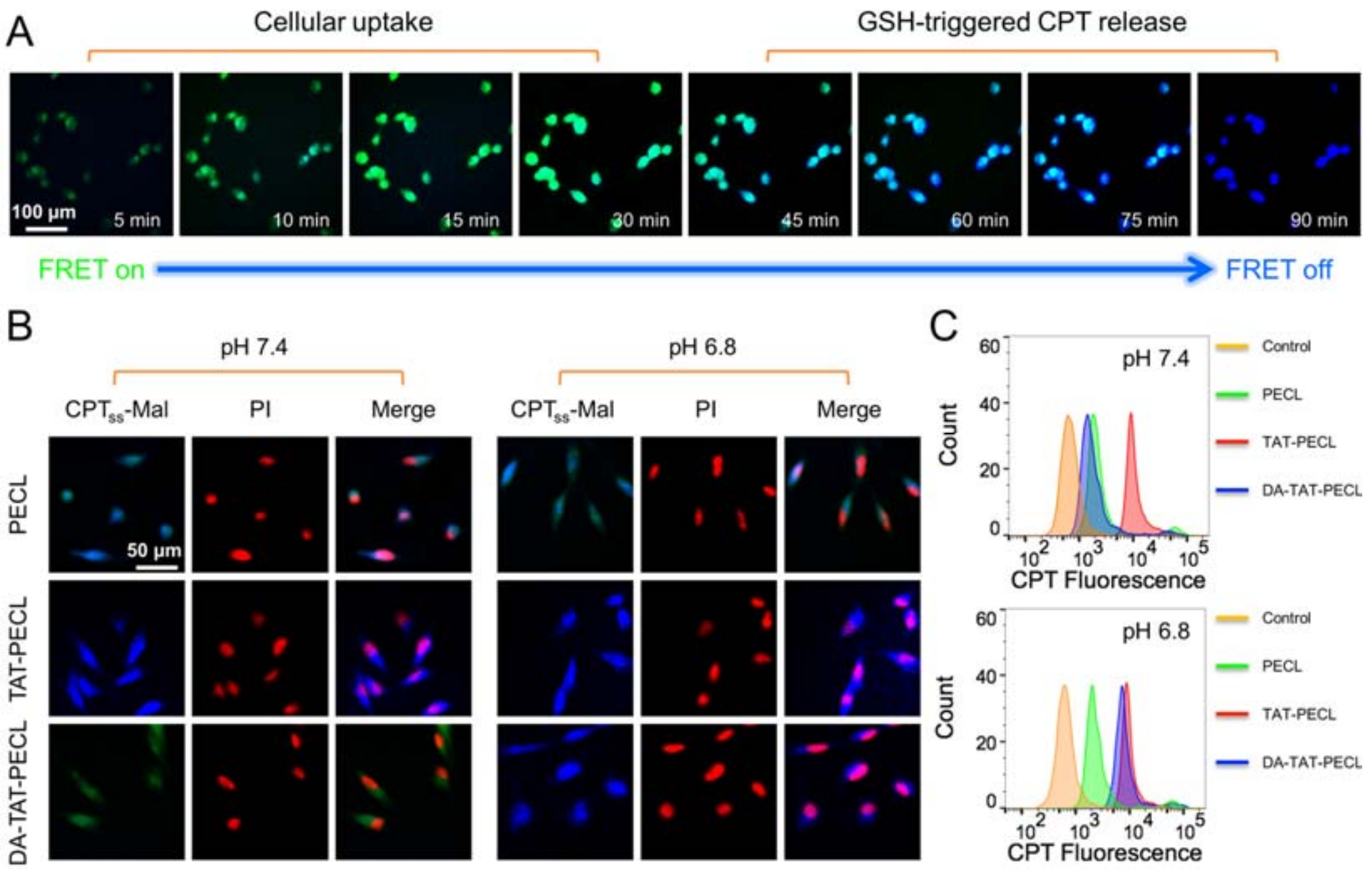

**Figure 6.** (A) Real-time fluorescence images in MDA-MB-231 cells showing the GSH-triggered FRET inactivation after treating with $(CPT)_2$-ss-Mal (CPT-equivalent dose: 1 mM). Green fluorescence (FRET on) indicated the uptake of $(CPT)_2$-ss-Mal by cells while blue fluorescence (FRET off) indicated the cleavage of maleimide thioether bond in $(CPT)_2$-ss-Mal and release of CPT from the drug dimer. (B) Fluorescence images and (C) flow cytometry analysis in MDA-MB-231 cells showing the pH-dependent cellular uptake. Cells were treated with $(CPT)_2$-ss-Mal-loaded PECL, TAT-PECL and DA-TAT-PECL micelles at pH 7.4 and pH 6.8 for 2 h (CPT-equivalent dose: 1 mM). Nuclei were stained with PI (red).

**pH-Dependent Cell Cytotoxicity.** We have shown that DA-TAT-PECL micelles could quickly deliver $(CPT)_2$-ss-Mal into MDA-MB-231 cells at pH 6.8, which were able to release CPT in response to intracellular GSH (**Figure 6**). In this case, cell cytotoxicity should also be pH-dependent for DA-TAT-PECL micelles. To demonstrate this, MDA-MB-231 cells were treated with various $(CPT)_2$-ss-Mal formulations at different pH values. Cell viability by alamarBlue assay indicated that $(CPT)_2$-ss-Mal-loaded DA-TAT-PECL micelles were more toxic at pH 6.8 relative to that at pH 7.4 (**Figure 7A**). This difference is attributed to the higher uptake of these micelles by MDA-MB-231 cells in acidic condition. There is no significant difference in the $IC_{50}$ values of $(CPT)_2$-ss-Mal at pH 7.4 and pH 6.8, and the same is true for the $(CPT)_2$-ss-Mal-loaded PECL and TAT-PECL micelles. Live/dead staining was performed to further evaluate the pH-dependent cell cytotoxicity, the result was consistent with the alamarBlue assay as more cells were dead if treated with $(CPT)_2$-ss-Mal-loaded DA-TAT-PECL micelles at pH 6.8 (**Figure 7B**).

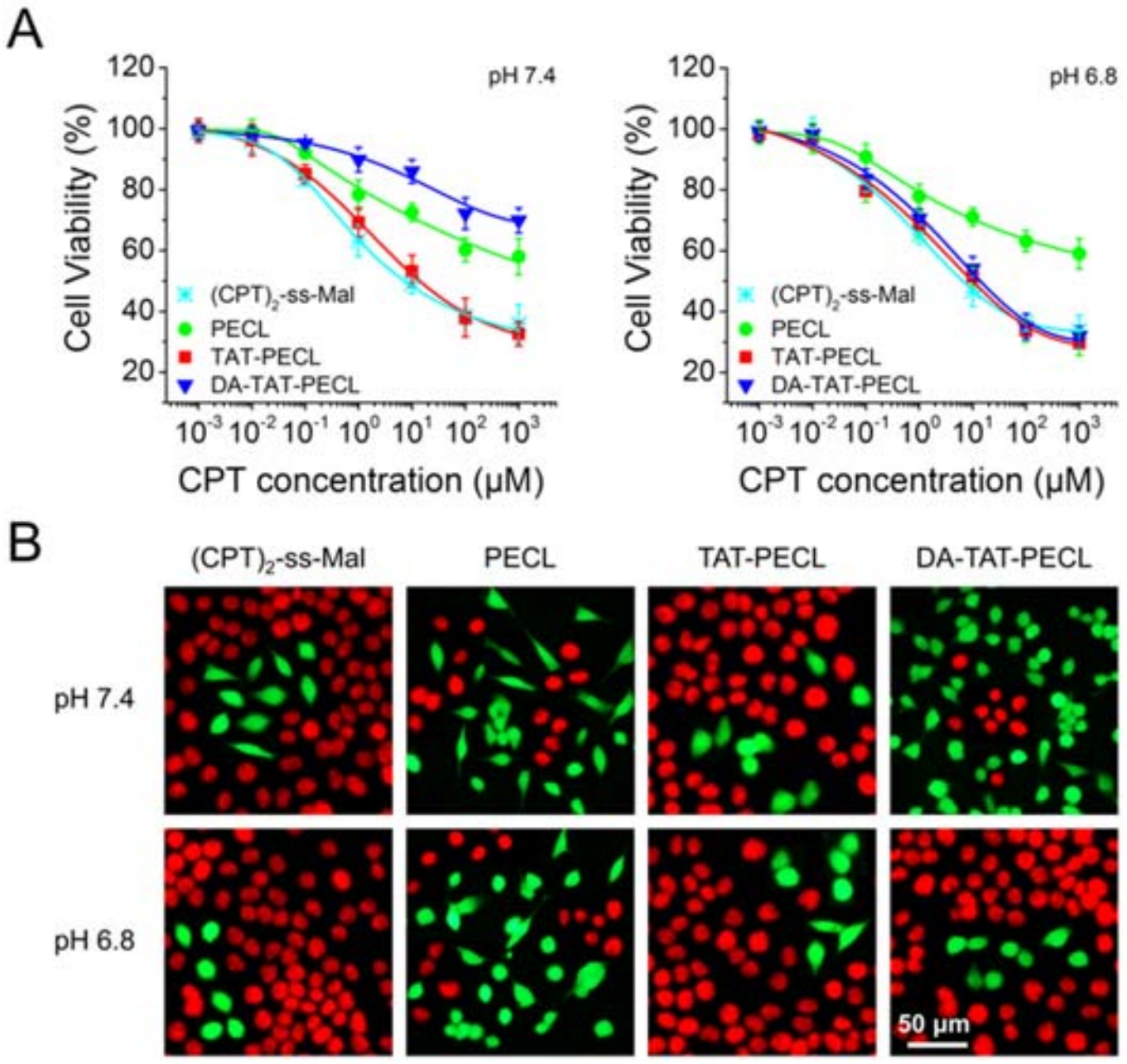

**Figure 7.** (A) Viability of MDA-MB-231 cells showing the pH-dependent cytotoxicity by alamarBlue assay. Cells were treated with different concentrations of $(CPT)_2$-ss-Mal and $(CPT)_2$-ss-Mal-loaded PECL, TAT-PECL and DA-TAT-PECL micelles at pH 7.4 and pH 6.8 for 48 h (n=5). (B) Fluorescence images showing the pH-dependent viability of MDA-MB-231 cells by calcein AM and propidium iodide double staining. Cells were treated with $(CPT)_2$-ss-Mal and $(CPT)_2$-ss-Mal-loaded PECL, TAT-PECL and DA-TAT-PECL micelles at pH 7.4 and pH 6.8 for 48 h (CPT-equivalent dose: 1 mM). The live cells were stained green while the dead cells were stained red.

**pH-Dependent Cell Apoptosis.** Annexin V-FITC/PI double-staining assay was used to evaluate the apoptosis in MDA-MB-231 cells induced by various $(CPT)_2$-ss-Mal formulations at different pH values. Compared to the treatment at pH 7.4, the percentage of MDA-MB-231 cells undergoing apoptosis after incubation with $(CPT)_2$-ss-Mal-loaded DA-TAT-PECL micelles at pH 6.8 was significantly increased (**Figure 8A** and **8B**). Similar to the cell cytotoxicity study, cells exposed to $(CPT)_2$-ss-Mal, $(CPT)_2$-ss-Mal-loaded PECL and TAT-PECL micelles showed pH-independent apoptosis characteristics. Therefore, it can be concluded that the DA-TAT-PECL system was able to induce more significant cell death and apoptosis upon acid stimuli.

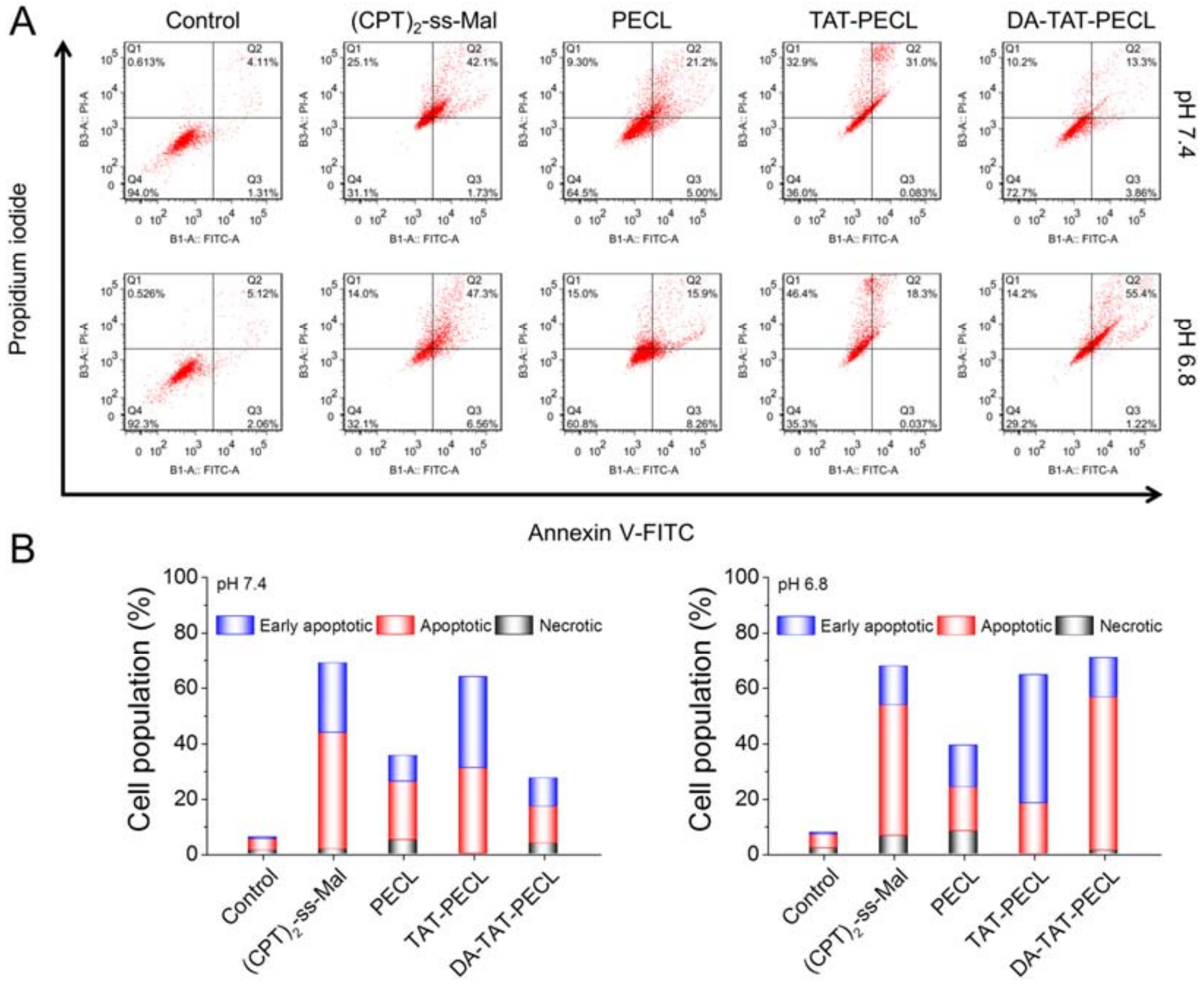

**Figure 8.** (A) Flow cytometry analysis in MDA-MB-231 cells showing the pH-dependent cell apoptosis by Annexin V-FITC/PI double-staining assay. Cells were treated with $(CPT)_2$-ss-Mal or $(CPT)_2$-ss-Mal-loaded PECL, TAT-PECL and DA-TAT-PECL micelles at pH 7.4 and 6.8 for 48 h (CPT-equivalent dose: 1 mM). X-axis represented FITC-labeled Annexin V positive cells and Y-axis represented PI positive cells. (B) Population of early apoptotic, apoptotic and necrotic MDA-MB-231 cells summarized from flow cytometry analysis.

## EXPERIMENTAL SECTION

**Materials.** TAT (CYGRKKRRQRRR-$NH_2$) was purchased from GenScript. Maleimide-PEG-OH ($M_n$ 5000), ε-caprolactone, 2,3-dimethylmaleic anhydride, campothecin, 2,3-dibromomaleimide, and tris(2-carboxyethyl)phosphine hydrochloride were purchased from Sigma-Aldrich. 2-Hydroxyethyl disulfide was purchased from Santa Cruz Biotechnology. AlamarBlue® and propidium iodide were purchased from ThermoFisher. Calcein AM and Annexin V-FITC apoptosis detection kit was purchased from eBioscience. All chemicals were purchased from commercial supplier and used as received.

**Amidization of TAT.** TAT (0.01 mmol, 16.6 mg) was dissolved in HEPES buffer (pH 8.5) and cooled to 0 °C, followed by adding 2,3-dimethylmaleic anhydride (DA) (0.03 mmol, 3.8 mg) and stirred for 24 h. The mixture was then dialyzed to remove the unreacted DA and lyophilize to get DA-TAT.

**Synthesis of Maleimide-PEG-PCL (Mal-PECL).** Mal-PECL was synthesized through ring-opening polymerization (ROP) of ε-caprolactone initiated by Maleimide-PEG-OH in the presence of stannous octoate ($Sn(Oct)_2$) as the catalyst. Briefly, Maleimide-PEG-OH (0.02 mmol, 100 mg), ε-caprolactone (1.1 mmol, 125 mg) and $Sn(Oct)_2$ (1 wt% of ε-caprolactone) were degassed under vacuum for 5 h with continuous stirring, then heated at 150 °C for 4 h. After that, the mixture was allowed to cool down, dissolved in dichloromethane, precipitated in cold ethanol, and collected by centrifugation. The Mal-PEG-PCL polymer was obtained by drying overnight under vacuum.

**Synthesis of TAT-PEG-PCL (TAT-PECL).** TAT-PECL was synthesized according to the literature.[24] Mal-PECL (0.005 mmol, 56.3 mg) was dissolved in acetonitrile, followed by rotary evaporation to form thin film at 37 °C and hydrated with phosphate buffer solution (PBS, pH 7.4). Then TAT (0.005 mmol, 8.3 mg) was added and stirred for 24 h. The unreacted TAT was removed by dialysis against PBS. The final product TAT-PEG-PCL was obtained by freeze drying.

**Synthesis of DA-TAT-PECL.** For the conjugation of DA-TAT, Mal-PECL (0.005 mmol, 56.3 mg) was dissolved in acetonitrile, followed by rotary evaporation to form thin film at 37 °C and hydrated with PBS buffer (pH 7.4). Then DA-TAT (0.005 mmol, 9.9 mg) was added and stirred for 24 h. The unreacted DA-TAT was removed by dialysis against PBS. The final product DA-TAT-PECL was obtained by freeze drying.

**Synthesis of $(CPT)_2$-ss.**[20] $(CPT)_2$-ss with a disulfide linker was synthesized through activating the hydroxyl group of CPT by triphosgene and then reacting with 2-hydroxyethyl disulfide. First, the dichloromethane solution of triphosgene (0.4 mmol, 118.7 mg) and 4-(dimethylamino)pyridine (DMAP, 1.0 mmol, 122.2 mg) was added dropwise into CPT (1.0 mmol, 348.4 mg) suspended in dichloromethane. Then, the mixture was stirred at room temperature for 6 h until the solution became clear, and reacted for another 24 h after addition of 2-hydroxyethyl disulfide (0.4 mmol, 61.7 mg). The product was obtained by removal of solvent through rotary evaporation and purified by silica gel column chromatography.

**Synthesis of $(CPT)_2$-ss-Mal.** The dithiomaleimide-based $(CPT)_2$-ss-Mal was synthesized by conjugating $(CPT)_2$-ss to 2,3-dibromomaleimide. Briefly, triethylamine (0.75 mmol, 75.9 mg) and TCEP.HCl (0.25 mmol, 71.6 mg) were dissolved in THF/methanol (1/3, v/v) and stired at room temperature for 1 h to remove the hydrochloric acid of TCEP.HCl. Then, the mixture was added dropwise into the solution of $(CPT)_2$-ss (0.25 mmol, 225.5 mg) and 2,3-dibromomaleimide (0.25 mmol, 63.7 mg), and stirred for another 24 h. The product was obtained by removal of solvent through rotary evaporation and purified by silica gel column chromatography.

**$^1$H NMR Determination.** $^1$H Nuclear magnetic resonance ($^1$H NMR) spectra were obtained on a Bruker Avance Ⅲ 500 apparatus. $CDCl_3$ or DMSO-$d_6$ was used as a solvent. Chemical shifts are expressed in parts per million, ppm ($\delta$).

**Preparation of Micelles.** Blank micelles were prepared by solvent evaporation method. Briefly, polymer was dissolved in 5 mL of tetrahydrofuran amd added dropwise into 10 mL of deionized water under high-speed stirring. Then the mixture was stired under a moderate speed to completely remove tetrahydrofuran. Drug-loaded micelles were prepared by dialysis method. Polymer and $(CPT)_2$-ss-Mal with a weight ratio of 1 : 1 were dissolved in 1 mL of DMSO and dropped into 9 mL of deionized water. After that, the mixture was transferred into a dialysis flask (MWCO 3500) against deionized water to remove the unloaded drug.

**DLS Determination.** Dynamic light scattering (DLS) (Malvern Zetasizer) was employed to detect the average size distribution of micelles. Each measurement was carried out in triplicate at 25 °C and an average value was reported.

**TEM Observation.** For transmission electron microscopy (TEM) observation, sample was prepared by negative staining. Briefly, the DA-TAT-PECL micellar solution was dropped onto a 300-mesh formvar film-supported copper grid for one minute and soaked up the excess solution by filter paper. Afterwards, a drop of phosphotungstic acid (2%, pH 7.0) was added onto the grid for another one minute and wicked off with filter paper. The grid was air dried and observed by JEOL JEM 1010 TEM operated at a voltage of 100 kV.

**Stability Against Protein Adsorption**. The stability of cationic charge-unshielded TAT-PECL micelles and shielded DA-TAT-PECL micelles were evaluated by size change against protein incubation. DLS was used to determine the size of the two micelles in the presence of 0.5 mg/mL bovine serum albumin (BSA) at different time points. Each measurement was carried out in triplicate at 25 °C and an average value was reported.

**Charge-Reversal Behavior**. TAT-PECL and DA-TAT-PECL polymers were dissolved in PBS at pH 7.4 or pH 6.8, and DLS was performed to measure their ζ-potentials at different time points. Each measurement was carried out in triplicate at 25 °C and an average value was reported.

**GSH-Induced FRET Inactivation**. A fluorescence spectrometer (FluoroMax-3, JY-Horiba) was used to record the emission spectra of $(CPT)_2$-ss-Mal solution in the presence of 10 mM GSH at different time points. The excitation wavelength was set at 370 nm. In the meantime, the pictures of $(CPT)_2$-ss-Mal solution after GSH incubation were taken under UV irradiation (365 nm).

**Drug Loading.** To detect the drug-loading content (LC) and encapsulation efficiency (EE), the $(CPT)_2$-ss-Mal-loaded micelles were lyophilized, redissolved in DMSO and measured by a Cary 50 UV-vis spectrophotometer (Agilent Technologies). The CPT concentration was quantified by absorpbance at 365 nm through standard curve. LC was calculated as LC(%) = weight of CPT in micelles/weight of CPT-loaded miclles, and EE was calculated as EE(%) = weight of CPT in micelles/weight of CPT in feed.

***In Vitro* Drug Release.** The lyophilized powder of CPT-loaded DA-TAT-PECL micelles were redissolved in 1 mL DMSO and dropped into 9 mL PBS (pH 7.4) with different concentrations of GSH (0, 10 μM and 10 mM). Then the micelles were dialyzed against the corresponding PBS solutions in a shaking incubator at 37 °C. The CPT concentrations in the media at selected time intervals from 0.5 to 48 h were measured by a FluoroMax-3 spectrofluorimeter upon the emission wavelength at 438 nm. The excitation wavelength was set at 370 nm.

**Cytocompatibility Assay.** Cytocompatibility of blank micelles was evaluated by alamarBlue assay and live/dead staining. For alamarBlue assay, MDA-MB-231 cells were seeded onto 48-well plates at a density of $1 \times 10^4$ cells/well for 24 h prior to treatment. Then PECL, TAT-PECL and DA-TAT-PECL blank micelles with concentrations in the range of 10 μg/mL to 500 μg/mL were added to the corresponding well. After 24 h incubation, cells were washed by PBS and treated with 300 μL of alamarBlue solution (10% alamarBlue, 80% media 199 (Gibcos), and 10% FBS, v/v) for further 3 h incubation. The plate was read on a Synergy HT fluorescent plate reader (BioTek Instruments Inc., Winooski, VT, USA) at an adsorption wavelength of 570 nm. Each experiment was performed in quintuplicate. Cell viability was calculated as cell viability(%) = $OD_{570}$(sample) − $OD_{570}$(blank)/$OD_{570}$(control) − $OD_{570}$(blank).

For live/dead staining, MDA-MB-231 cells were seeded onto 12-well plates at a density of $2 \times 10^4$ cells/well. After 24 h incubation with different concentrations of blank micelles, cells were treated with 10 μM Calcein-AM for 10 minutes and 1 μM propidium iodide (PI) for 10 min. The images were taken using an Olympus BX 50 Fluorescence Microscope. The Calcein-AM fluorescence (green) indicated live cells and the PI fluorescence (red) indicated dead cells.

**Intracellular FRET Inactivation.** To observe the GSH-triggered intracellular FRET inactivation, MDA-MB-231 breast cancer cells ($10 \times 10^4$ cells/well) grown in 6-well plates were treated with $(CPT)_2$-ss-Mal at a CPT dose of 1 mM. The images were taken at different time points under the exitation wavelength of 365 nm using an Olympus BX 50 Fluorescence Microscope. The Mal fluorescence (green) indicated FRET activation and the CPT fluorescence (blue) indicated FRET inactivation.

**Cellular Uptake Observed by Fluorescence Microscopy.** MDA-MB-231 cells were seeded onto 6-well plates at a density of $10 \times 10^4$ cells/well and allowed to grow for 24 h. Then the culture medium was replaced with $(CPT)_2$-ss-Mal-loaded PECL, TAT-PECL and DA-TAT-PECL micelles suspended in neutral medium (pH 7.4) and acidic medium (pH 6.8) (at a CPT dose of 1 mM). After 2 h of incubation, cells were washed with PBS, fixed by paraformaldehyde, stained with propidium iodide (PI), and imaged by fluorescence microscopy. CPT (blue) and PI (red) were excited at 365 nm and 535 nm, respectively.

**Cellular Uptake Measured by Flow Cytometry.** MDA-MB-231 cells were seeded onto 6-well plates at a density of $10 \times 10^4$ cells/well and allowed to grow for 24 h. The culture medium were replaced with $(CPT)_2$-ss-Mal-loaded PECL, TAT-PECL and DA-TAT-PECL micelles suspended in neutral medium (pH 7.4) and acidic medium (pH 6.8) (at a CPT dose of 1 mM). After 2 h incubation, cells were washed with PBS, harvested by trypsin, collected in tubes and centrifuged at 1500 rpm for 5 min, and then resuspended in PBS. Miltenyi Biotec 8-color MACSQuant VYB flow cytometer was used to analyze the intracellular fluorescence intensity.

**Cell Cytotoxicity.** The cytotoxicity of various CPT formulations against MDA-MB-231

cells at different pH values was evaluated by alamarBlue assay and live/dead staining. For alamarBlue assay, 1 $\times 10^4$ cells were seeded in 48-well plates for 24 h, and the culture medium were replaced with free $(CPT)_2$-ss-Mal and $(CPT)_2$-ss-Mal-loaded PECL, TAT-PECL and DA-TAT-PECL micelles suspended in neutral medium (pH 7.4) and acidic medium (pH 6.8). CPT doses were ranging from $10^{-3}$ μM to $10^3$ μM. 24 h later, cells were washed with PBS, incubated with 300 μL of alamarBlue solution for further 3 h, and read on a Synergy HT fluorescent plate reader.

For live/dead staining, MDA-MB-231 cells were seeded onto 12-well plates at a density of $2 \times 10^4$ cells/well, and incubated with various CPT formulations at different pH conditons for 24 h (at a CPT dose of 1 mM). Then cells were treated with 10 μM Calcein-AM for 10 minutes and 1 μM propidium iodide (PI) for 10 min. The images were taken using an Olympus BX 50 Fluorescence Microscope. The Calcein-AM fluorescence (green) indicated live cells and the PI fluorescence (red) indicated dead cells.

**Cell Apoptosis and Necrosis.** Cell apoptosis and necrosis induced by various CPT formulations at different pH values were analyzed by flow cytometry using the Annexin V-FITC/PI apoptosis detection kit. MDA-MB-231 cells seeded at $10 \times 10^4$ cells/well were incubated with free $(CPT)_2$-ss-Mal and $(CPT)_2$-ss-Mal-loaded PECL, TAT-PECL and DA-TAT-PECL micelles suspended in neutral medium (pH 7.4) and acidic medium (pH 6.8) (at a CPT dose of 1 mM). After 24 h of incubation, cells were trypsinized, collected, and labeled with Annexin V-FITC and PI according to the manufacturer's protocols prior to fluorescence analysis on a flow cytometer.

## CONCLUSIONS

In summary, we developed an acid-active tumor targeting DA-TAT-PECL nanoplatform loaded with reduction-sensitive $(CPT)_2$-ss-Mal. Shielding of the cationic charges of TAT with DA effectively inhibited its nonspecific interactions with proteins in blood circulation. Meanwhile, the original TAT is regenerated in the acidic tumor microenvironment, facilitating the cellular entry of the nanoparticles via cell-penetrating peptide-mediated endocytosis and electrostatic interaction. Once the nanoparticles were internalized by tumor cells and escaped from the endo/lysosomes, CPT would be released from $(CPT)_2$-ss-Mal in response to the high concentration of intracellular GSH. The change in FRET signals between CPT and maleimide thioether bond was able to report the drug release process. This nanomedicine will provide a promising platform for effective cancer therapy.

## AUTHOR INFORMATION

**Corresponding Author**

*E-mail: shaobingzhou@home.swjtu.edu.cn; Zi.Chen@dartmouth.edu

**Author Contributions**

⊥These authors contributed equally.

**Notes**

The authors declare no competing financial interest.

## ACKNOWLEDGMENTS

This work was partially supported by the Society in Science Branco Weiss Fellowship, administered by ETH Zürich. The research reported in this paper was in part supported by the National Cancer Institute of the National Institutes of Health under Award (No. U01CA202123) and the National Natural Science Foundation of China (No. 51603172).